\documentclass[11pt]{article}

\usepackage{amssymb}
\usepackage[dvips]{graphicx}
\usepackage{citesort}

\def\bra#1{\left\langle#1\right|}
\def\ket#1{\left|#1\right\rangle}

\begin{document}
\vspace*{1cm}
\begin{center}
{\LARGE\bf\sf Recursive representation of the torus 1-point conformal block} \\
\end{center}

\begin{center}

\vspace*{2cm}

    {\large\bf\sf
    Leszek Hadasz${}^\dag$\footnote{\emph{e-mail}: hadasz@th.if.uj.edu.pl}$\!\!\!\!,\ \,$
    Zbigniew Jask\'{o}lski${}^\ddag$\footnote{\emph{e-mail}: jask@ift.uni.wroc.pl}
    and
    Paulina Suchanek${}^\ddag$\footnote{\emph{e-mail}: paulina@ift.uni.wroc.pl}
    }
     \\
\vskip 3mm
    ${}^\dag$ M. Smoluchowski Institute of Physics,
    Jagiellonian University \\
    Reymonta 4,
    30-059~Krak\'ow, Poland, \\

\vskip 3mm
    ${}^\ddag$ Institute of Theoretical Physics,
    University of Wroc{\l}aw \\
    pl. M. Borna, 50-204~Wroc{\l}aw, Poland. \\
\end{center}

\vspace*{1cm}
\begin{abstract}
The recursive relation for the 1-point conformal block on a torus is derived
and used to prove the identities between conformal blocks
recently conjectured by Poghossian in \cite{Poghossian:2009mk}.
As an illustration of the efficiency of the recurrence  method
the modular invariance of the 1-point Liouville correlation function is numerically
analyzed.
\end{abstract}

\vspace*{\fill}


\section{Introduction}

An interesting duality between certain class of N = 2
superconformal field theories in four dimensions and the
two-dimensional Liouville field theory has been recently
discovered by Alday, Gaiotto and Tachikawa \cite{Alday:2009aq}. An
essential part of this relation  actively studied in a number of
papers
\cite{Wyllard:2009hg,Marshakov:2009gs,Gaiotto:2009ma,Mironov:2009dr,Alday:2009fs,Marshakov:2009kj,Mironov:2009qn,Bonelli:2009zp,Alba:2009fp}
is an exact correspondence between  instanton parts of  Nekrasov
partition functions in N=2 SCFT and  conformal blocks of the
2-dimensional CFT. This in particular concerns two basic objects
of CFT: the 4-point conformal block on a sphere and the 1-point
conformal block on a torus. These special cases has been recently
analyzed by Poghossian \cite{Poghossian:2009mk} who applied the
recursive relation for the 4-point blocks on a sphere discovered
long time ago by Alexei Zamolodchikov in CFT
\cite{Zamolodchikov:ie,Zamolodchikov:2,Zamolodchikov:3} to the
instanton part of the Nekrasov partition function with four
fundamentals. He also proposed and verified to certain order a
recursive relation for the Nekrasov function with one  adjoint
hypermultiplet. It was observed that by the AGT duality this
yields previously unknown recursive relation for conformal blocks
in 2-dimensional CFT.

The  aim of the present paper is to provide a complete derivation
of the recursive relation for 1-point conformal blocks on a torus
conjectured by Poghossian \cite{Poghossian:2009mk}. Our method is
based on the algebraic properties of the 3-point conformal blocks
and the analytic structure of the inverse of the Gram matrices in
Verma modules \cite{Hadasz:2006sb}. It parallels to large extend
Zamolodchikov's original derivation for the 4-point blocks
\cite{Zamolodchikov:ie,Zamolodchikov:2,Zamolodchikov:3}. We set
our notation in Section 2 and present the derivation in Section~3.

As an application of the recurrence relations we prove two
important identities between conformal blocks conjectured  by
Poghossian  \cite{Poghossian:2009mk}. The first one relates the
1-point block on a torus with certain 4-point elliptic blocks on a
sphere. The second relates 4-point elliptic blocks for different
values of central charges, external weights and the elliptic
variable $q$. It had been motivated by the intriguing relation
 between the Liouville 1-point  correlation function on the torus and the
Liouville 4-point
 correlation function on the sphere  recently proposed
by Fateev, Litvinov, Neveu and Onofri \cite{Fateev:2009me}. The
derivation of both relations is presented in Section 4. They in
particular  imply\footnote{See the next section for our notation
conventions.}:
\begin{eqnarray}
\label{rel:among:blocksI}
\mathcal{H}_{c,
\Delta_\alpha}^{\lambda}\left(q^2\right)
&=&
\mathcal{H}_{c',{\Delta'}_{\!\!\alpha'}}
\!\left[^{{1\over 2b'}\;\;{\lambda\over \sqrt{2}}}_{{{1\over 2b'}\;\;\:{1\over 2b'}}}\right]\!(q)
\;,\;\;\;b'=\textstyle {b\over
\sqrt{2}}\;,\;\;\;\alpha' = {\sqrt{2}\alpha  }\ ,
\end{eqnarray}
and
\begin{eqnarray}
\label{rel:among:blocksII}
\mathcal{H}_{c,
\Delta_\alpha}^{\lambda}\left(q^2\right)
&=&
\mathcal{H}_{c',{\Delta'}_{\!\!\alpha'}}
\!\left[^{{b'\over 2}\;\;{\lambda\over \sqrt{2}}}_{{{b'\over 2}\;\;\:{b'\over 2}}}\right]\!(q)
\;,\;\;\;b'= \textstyle
{\sqrt{2}b}\;,\;\;\;\alpha' = {\sqrt{2}\alpha  }\ .
\end{eqnarray}
These identities, which can be seen as chiral versions of the identity
for the Liouville correlation functions proposed in
\cite{Fateev:2009me}, constitute the most interesting
results of the present work. They  provide for instance  a new tool for analyzing
virtually untouched problem of modular invariance of nontrivial
torus 1-point functions in CFT. We hope to report on some of they
consequences in the forthcoming paper.

Beside  theoretical applications the recursive relation for
1-point block on a torus gives a very efficient numerical
method of analyzing the modular bootstrap in CFT. As an
illustration of this method we present in Section 5 some numerical
checks of modular bootstrap in the Liouville theory.

\section{1-point functions}
Consider a primary field $\phi_{\lambda,\bar\lambda}$
with the conformal weights $(\Delta_{\lambda} ,\Delta_{\bar\lambda})$ for which the following parametrization is assumed:
$$
\Delta_\lambda = \frac14\left(Q^2 - \lambda^2\right)
 \;,\;\;\;c=1+6Q^2\;,\;\;\;Q=b+b^{-1}.
$$
In terms of the CFT on the complex plane the 1-point correlation function
of $\phi_{\lambda,\bar\lambda}$ on a torus takes the form:
\begin{eqnarray*}
\langle \phi_{\lambda,\bar \lambda} \rangle  = \mathrm{Tr}\left( e^{- (Im \tau) \hat{H} + i (Re \tau) \hat{P}} \phi_\lambda(1,1)
\right)
= (q \bar q)^{-\frac{c}{24}} \mathrm{Tr} \left( q^{L_0} \bar q^{\bar L_0} \phi_\lambda(1,1)\right),
\end{eqnarray*}
where
$\tau$ is the torus modular parameter and
\begin{eqnarray*}
\hat{H} &=& 2 \pi (L_0 + \bar L_0) - \frac{\pi c}{6}, \qquad
\hat{P} = 2\pi (L_0 - \bar L_0), \qquad
q = e^{2 \pi i \tau}.
\end{eqnarray*}
The trace can be calculated in the standard bases of Verma modules:
$$
\nu_{\Delta,M}
 = L_{-M}\nu_{\Delta} \; \equiv \;
L_{-m_j}\ldots L_{-m_1}\nu_{\Delta}\,,
$$
where $M = \{m_1,m_2,\ldots,m_j\}\subset \mathbb{N}$ stands for an arbitrary ordered  set of  indices
$
m_j \leq \ldots \leq m_2 \leq m_1,
$
and $\nu_{\Delta}\in \mathcal{V}_{\Delta}$ is the highest weight state.
This yields:
\begin{eqnarray}
\label{coorfunct:1}
\langle \phi_{\lambda,\bar \lambda} \rangle &=& (q \bar q)^{-\frac{c}{24}}
\sum\limits_{(\Delta,\bar\Delta)}\; \sum\limits_{n=0}^\infty \;
 q^{\Delta + n} \, \bar{q}^{\bar \Delta + n}   \\
 \nonumber
 &\times &
 \sum\limits_{\begin{array}{c}\scriptstyle {n=|M|=|N|}\\[-6pt]\scriptstyle {n=|\bar M|=|\bar N|}
\end{array}}
\Big[ B^n_{c,\Delta} \Big]^{MN}
\Big[\bar B^n_{c,\bar \Delta} \Big]^{\bar M \bar N}
\bra{\nu_{\Delta,M} \otimes \bar{\nu}_{\bar \Delta,\bar M}}  \phi_{\lambda,\bar \lambda}(1,1) | \nu_{\Delta,N} \otimes\bar{\nu}_{\bar\Delta,\bar N}  \rangle
\end{eqnarray}
where the sum over all weights $(\Delta,\bar\Delta)$ from the spectrum of the theory is assumed,
$ |M|= m_1+\dots+m_j$
and
{\small $\left[B^{n}_{c, \Delta}\right]^{MN}$} is the
inverse of the Gram matrix
\[
\left[B^{n}_{c, \Delta}\right]_{MN}
=
\left\langle\nu_{\Delta,N}\big|\nu_{\Delta,M}\right\rangle,
\hskip 1cm
|M| = |N| = n,
\]
at the level $n$.

The 3-point correlation function  in (\ref{coorfunct:1}) can be
 expressed as a product of the 3-point (``left'' and ``right'') conformal blocks and a structure constant:
\begin{eqnarray}
\label{three:point:function}
\nonumber
 \bra{\nu_{\Delta,M} \otimes \bar{\nu}_{\bar\Delta,\bar M}}
 \phi_{\lambda,\bar \lambda}(1,1) | \nu_{\Delta,N} \otimes\bar{\nu}_{\bar\Delta,\bar N}  \rangle
 & = &
  \rho(\nu_{\Delta,N}, \nu_{\lambda} , \nu_{\Delta,M}) \, \rho( \nu_{\bar\Delta,\bar N},   \nu_{\bar\lambda} ,  \nu_{\bar\Delta,\bar M})\,
  C^{\lambda,\bar\lambda}_{\Delta,\bar\Delta},
  \\[0pt]
  \nonumber
  \\[-10pt]
  \nonumber
 && \hspace{-160pt}
 C^{\lambda,\bar\lambda}_{\Delta,\bar\Delta}
\;=\;
\bra{\nu_{\Delta} \otimes{\nu}_{ \bar\Delta}}  \phi_{\lambda,\bar \lambda}(1,1) | \nu_{\Delta} \otimes{\nu}_{\bar\Delta}  \rangle.
\end{eqnarray}
Introducing the 1-point conformal block:
  \begin{eqnarray}
  \nonumber
\mathcal{F}_{c, \Delta}^{\lambda}(q)
    &=& q^{\Delta -\frac{c}{24}} \,
\sum_{n =0}^{\infty} q^{n} \,
    F^{\lambda, n}_{c, \Delta},
\\[-8pt]
\label{blockCoeff}
\\[-4pt]
\nonumber
F^{\lambda, n}_{c, \Delta} &=&
\sum_{n=|M|=|N|}  \rho(\nu_{\Delta,N}, \nu_{\lambda} , \nu_{\Delta,M})
\left[ B^n_{c,\Delta} \right]^{MN},
 \end{eqnarray}
one can write  the 1-point correlation function
on the torus
in the following form:
 $$
 \langle \phi_{\lambda,\bar \lambda} \rangle = \sum_{(\Delta,\bar\Delta)} \mathcal{F}_{c, \Delta}^{\lambda}(q)\,
  \mathcal{F}_{c, \bar \Delta}^{\lambda}(\bar q) \, C^{\lambda,\bar\lambda}_{\Delta,\bar\Delta}\ .
 $$

\section{Recursive relations}
For arbitrary vectors $\xi_i\in {\cal V}_{\Delta_i}$ the 3-point block
$
\rho(\xi_3, \xi_2 , \xi_1)
$
is a polynomial function of the weights $\Delta_i,$
completely determined by the conformal Ward identities \cite{Hadasz:2006sb} and
the normalization condition
$ \rho(\nu_3, \nu_2 , \nu_1) = 1$.
It follows from (\ref{blockCoeff}) that
the block coefficients
\(
F^{\lambda, n}_{c, \Delta}
\)
are polynomials in the external weight $\Delta_\lambda$ and
rational functions of the central charge $c$ and the intermediate weight $\Delta$
with the locations of poles
determined by the zeroes of the determinant of the Gram matrix
$\left[ B^{\,{n}}_{c,\Delta}\right]_{M,N}.$

In the generic case the Verma module $ {\cal V}_{\Delta_{rs}+rs}$ is not reducible
and the only singularities  of $\left[ B^{\,{n}}_{c,\Delta}\right]^{M,N}$  as a function of
 $\Delta$ are simple poles at
\begin{eqnarray}
\nonumber
\label{delta:rs}
\Delta_{rs}(c)
& = & \frac{Q^2}{4}- \frac14 \left(r b + {s\over b}\right )^2
\end{eqnarray}
where $r, s\in\mathbb{Z},\;\;\;r\geq 1,\;s\geq 1,\;\;\;1\leq rs\leq n$.

As a function of $c$ the inverse Gram matrix at the level $n$ has simple poles at the locations
\begin{eqnarray}
\nonumber
\label{zero:rs}
 c_{rs}(\Delta)
&=& 1 + 6\left(b_{rs}(\Delta) +\frac{1}{b_{rs}(\Delta)}\right)^2,
\\
\nonumber
b^2_{rs}(\Delta) &=& \frac{1}{ 1- r^2} \, \left(rs -1 + 2\,\Delta  +
    {\sqrt{{\left( r - s \right) }^2 +
        4\,\left(r\,s -1 \right) \,\Delta  +
        4\,{\Delta }^2}} \right).
\end{eqnarray}
where $r, s\in\mathbb{Z},\;\;\;r\geq 2,\;s\geq 1,\;\;\;1\leq rs\leq n$.

The block's coefficient $ F^{\lambda, n}_{c, \Delta}$ can  be expressed either as
a sum over the poles in the intermediate weight:
\begin{equation}
\label{first:expansion:Delta}
F^{\lambda, n}_{c, \Delta}
\; = \;
{\rm h}^{\lambda,n}_{c}
+
\begin{array}[t]{c}
{\displaystyle\sum} \\[-6pt]
{\scriptscriptstyle
1 \leq rs \leq n}
\end{array}
\frac{
{\mathcal R}^{\lambda,n}_{c,\,rs}\!
}
{
\Delta-\Delta_{rs}(c)
}\,,
\end{equation}
or in the central charge:
 \begin{equation}
\label{first:expansion:c}
F^{\lambda, n}_{c, \Delta}
\; = \;
{\rm f}^{\lambda, n}_{\Delta}
+
\begin{array}[t]{c}
{\displaystyle\sum} \\[-6pt]
{\scriptscriptstyle
1 < rs \leq n}
\end{array}
\frac{
\widetilde{{\mathcal R}}^{\lambda, n}_{\Delta,\,rs}
}
{
c-c_{rs}(\Delta)
}\,.
\end{equation}
 The residues at $\Delta_{rs}(c)$ and $c_{rs}(\Delta)$ are simply related:
\begin{eqnarray}
\nonumber
\label{c:Delta:relation}
\widetilde{\mathcal R}^{\lambda,n}_{\Delta,\,rs}
&=&-
{\partial c_{rs}(\Delta)\over \partial \Delta}\,
{\mathcal R}^{\lambda,n}_{c_{rs}(\Delta),\,rs}
\\[6pt] \nonumber
\label{der}
{\partial c_{rs}(\Delta)\over \partial \Delta}
&=& 4 \, \frac{c_{rs}(\Delta)-1}{\left(r^2-1\right)(b_{rs}(\Delta))^4 - \left(s^2-1\right)}.
\end{eqnarray}
In order to calculate the residue
at $\Delta=\Delta_{rs}$
it is useful to choose a specific basis
in the Verma module ${\cal V}_{{\Delta}}.$
Let us introduce the state:
\begin{eqnarray*}
\chi_{rs}^{\Delta} = \sum_{|M|=rs} \chi_{rs}^M L_{-M} \nu_{\Delta}
\end{eqnarray*}
where
$\chi_{rs}^M $ are the coefficients of the singular vector $\chi_{rs} $
in the standard basis of ${\cal V}_{\Delta_{rs}}:$
 $$
  \chi_{rs} = \sum_{|M|=rs} \chi_{rs}^M L_{-M} \nu_{\Delta_{rs}}.
  $$
The family of states $\left\{L_{-N} \chi_{rs}^{\Delta}\right\}_{|N|=n-rs}$   can be completed to a full basis in the
Verma module ${\cal V}_{\Delta}$
at the level $n>rs$.
Working in this basis one gets \cite{Hadasz:2006sb}:
\begin{equation}
\nonumber
\label{res:1a}
{\mathcal R}^{\lambda, n}_{c,\,rs}
=
\lim_{\Delta\to\Delta_{rs}} (\Delta - \Delta_{rs}(c)) \,
F^{\lambda, n}_{c, \Delta}
 \; = \;
 A_{rs}(c)
\hskip -10pt
\sum_{n-rs=|M|=|N|}
\hskip -10pt
    \rho(L_{-N}\chi_{rs}, \nu_{\lambda} , L_{-M}\chi_{rs})
   \ \left[B^{{n}-rs}_{c, \Delta_{rs}+rs}\right]^{MN},
\end{equation}
where
$$
A_{rs}(c)
\; = \;
\lim_{\Delta\to\Delta_{rs}}
\left(\frac{\left\langle\chi_{rs}^{\Delta}|\chi_{rs}^{\Delta}\right\rangle}{\Delta - \Delta_{rs}(c)}
\right)^{-1}.
$$
It is convenient to normalize the singular vector $\chi_{rs}$ such that the coefficient in front of $(L_{-1})^{rs}$  equals~1. For this normalization
the exact form of the coefficient $A_{rs}(c)$  was first proposed by Al.~Zamolodchikov in \cite{Zamolodchikov:ie} and
then justified  in
\cite{Zamolodchikov:2003yb}. It reads:
\begin{eqnarray*}
A_{rs}(c)
\; = \;
\frac12
\hspace{-35pt}\prod_{\begin{array}{c}\\[-24.5pt]\scriptstyle p=1-r\\[-6pt]\hspace{30pt}\scriptstyle (p,q)\neq (0,0),(r,s)
\end{array}}^{r}
\hspace{-35pt}\prod_{q=1-s}^{s}
\left( pb +\frac{q}{b}\right)^{-1}.
\end{eqnarray*}
Using the
factorization formula:
\begin{eqnarray*}
\rho(L_{-N}\chi_{rs} , \nu_{\lambda} , L_{-M}\chi_{rs} ) \, = \,
\rho(L_{-N}\nu_{\Delta_{rs}+rs}, \nu_{\lambda} , L_{-M}\nu_{\Delta_{rs}+rs}) \
\rho(\chi_{rs}, \nu_{\lambda} , \chi_{rs})
\end{eqnarray*}
one can show
that the residue   is proportional to the lower order block coefficient:
\begin{eqnarray}
\nonumber
\label{res:1b}
&& \hskip -5mm
{\mathcal R}^{\lambda, n}_{c,\,rs}
 \; = \;  A_{rs}(c) \, \rho(\chi_{rs} , \nu_{\lambda} , \chi_{rs} ) \, F^{\lambda, n-rs}_{c, \Delta_{rs}+rs}.
\end{eqnarray}
For the normalized singular vector  $\chi_{rs}$ one gets \cite{Hadasz:2006sb}:
\begin{eqnarray*}
\rho(\chi_{rs} , \nu_{\lambda} , \chi_{rs} )
&=&
\rho(\chi_{rs}, \nu_{\lambda} , \nu_{\Delta_{rs}+rs}) \
\rho(\nu_{\Delta_{rs}}, \nu_{\lambda} , \chi_{rs})
\;\ = \;\
P^{rs}_{c}\!\left[^{\hskip 10pt \Delta_{\lambda}}_{\Delta_{rs}+rs}\right] \,
P^{rs}_{c}\!\left[^{\Delta_{\lambda}}_{\Delta_{rs}}\right]
\end{eqnarray*}
where the fusion polynomials are  defined by:
\begin{eqnarray}
\nonumber
\label{fusionPoly}
P^{rs}_{c}\!\left[^{\Delta_2}_{\Delta_1}\right]
\;=\!\!\!\!\!\!
\prod\limits_{\begin{array}{c} \\[-22pt]\scriptstyle p=1-r\\[-6pt]\scriptstyle p+r=1\,  {\rm mod}\, 2
\end{array}}^{r-1}
\!\!\!
\prod\limits_{\begin{array}{c}\\[-22pt]\scriptstyle q=1-s\\[-6pt]\scriptstyle q+s=1\, {\rm  mod}\, 2
\end{array}}^{s-1}
\!\!\!\!\!\!\!\!\!
\left(\frac{\lambda_2+\lambda_1 +p b + q b^{-1}}{2}\right) \left(\frac{\lambda_2-\lambda_1 +p b + q b^{-1}}{2}\right)
\end{eqnarray}
and
$
\Delta_i = \frac14\left(Q^2 - \lambda_i^2\right).
$
In the case under consideration:
\begin{eqnarray}
\label{rho:explicit}
P^{rs}_{c}\!\left[^{\hskip 10pt \Delta_{\lambda}}_{\Delta_{rs}+rs}\right] \,
P^{rs}_{c}\!\left[^{\Delta_{\lambda}}_{\Delta_{rs}}\right]&=&
\prod_{\begin{array}{c}\\[-22pt]\scriptstyle k=1\\[-6pt]\scriptstyle k=1\, {\rm  mod}\, 2
\end{array}}^{2r-1}
\prod_{\begin{array}{c}\\[-22pt]\scriptstyle l=1\\[-6pt]\scriptstyle l=1\, {\rm  mod}\, 2
\end{array}}^{2s-1}
\left(\frac{\lambda + k b + l b^{-1}}{2}\right)
\\[6pt]
\nonumber
& \times&
\left(\frac{\lambda - k b - l b^{-1}}{2}\right)
\left(\frac{\lambda + k b - l b^{-1}}{2}\right)
\left(\frac{\lambda - k b + l b^{-1}}{2}\right).
\end{eqnarray}
The last step in our derivation  is to find the non-singular terms in
the equations (\ref{first:expansion:Delta}), (\ref{first:expansion:c}).
Let us start with the expansion (\ref{first:expansion:c}).
Since ${\rm f}^{\lambda, n}_{\Delta}$  does not depend on the central charge it can be calculated
from the $c\to \infty$ limit:
$$
\sum_{n =0}^{\infty} q^{n} {\rm f}^{\lambda, n}_{\Delta} =
\lim_{c\to \infty}q^{-\Delta+\frac{c}{24}} \mathcal{F}_{c, \Delta}^{\lambda}(q).
$$
Note that
 the block's coefficients (\ref{blockCoeff}) depend on $c$ only via the inverse
Gram matrix $\left[B^{n}_{c, \Delta}\right]^{M,N}$.
Analyzing the polynomial dependence of the Gram matrix minors on $c$
one can show \cite{Hadasz:2006sb} that the only element of the inverse Gram matrix which  does not vanish in the limit
  $c \to \infty$ is the diagonal one corresponding to the state $L_{-1}^n \ket{\nu_\Delta}$:
\begin{eqnarray*}
\lim_{c \to \infty}\Big[ B_{c,\Delta}^n\Big]^{1\hspace{-1mm}\mathbb{I}\, 1\hspace{-1mm}\mathbb{I}}
 = \frac{1}{\bra{\nu_\Delta} L_{1}^n  L_{-1}^n \ket{\nu_\Delta}}
= \frac{1}{n!\big(2\Delta\big)_n},
\end{eqnarray*}
where
$
\big(a\big)_n = \frac{\Gamma(a+n)}{\Gamma(a)}
$
is the Pochhammer symbol.
The 3-point block for this state is given by:
$$
\rho(L_{-1}^n \nu_{\Delta}, \nu_{\lambda}, L_{-1}^n \nu_{\Delta}) =
\sum_{k=0}^n
\frac{1}{(n-k)!}
\left(\frac{n!}{k!}\right)^2\,\frac{\Gamma(2\Delta + n)}{\Gamma(2\Delta + k)}\,
\frac{\Gamma(\Delta_{\lambda}+k)}{\Gamma(\Delta_{\lambda}-k)}.
$$
Thus the $c\to \infty $ limit of the 1-point block reads:
$$
\lim_{c\to \infty} q^{-\Delta+\frac{c}{24}}\mathcal{F}_{c, \Delta}^{\lambda}(q) =
\sum_{n=0}^{\infty} q^n  \sum_{k=0}^n
\frac{n!}{(n-k)!(k!)^2}\,\frac{\Gamma(2\Delta)}{\Gamma(2\Delta + k)}\,
\frac{\Gamma(\Delta_{\lambda}+k)}{\Gamma(\Delta_{\lambda}-k)},
$$
which yields:
\[
\nonumber
\label{cInfty}
\sum_{n =0}^{\infty} q^{n} {\rm f}^{\lambda, n}_{\Delta}
=
 \frac{1}{1-q}\ {}_2F_1\!\left(\Delta_{\lambda},1-\Delta_{\lambda};2\Delta;\frac{q}{q-1}\right).
\]

The $\Delta\to\infty$ asymptotics of the block is even easier to obtain  and leads to a more convenient recursion.
With the help of the Ward identities  one easily shows that
for $\Delta\to\infty:$
\[
\nonumber
\label{rho:at:large:delta}
\rho(\nu_{\Delta,N}, \nu_{\lambda} , \nu_{\Delta,M})
=
\rho(\nu_{\Delta,N}, \nu_{Q} , \nu_{\Delta,M})\left(1+{\cal O}(\Delta^{-1})\right)
\]
where $\nu_{Q}$ is the highest weight state in the vacuum Verma module ($\Delta_{{Q}} = 0$).
Since (for $|M| = |N| = n$):
\[
\rho(\nu_{\Delta,N}, \nu_{Q} , \nu_{\Delta,M})  = \left[ B^{\,{n}}_{c,\Delta}\right]_{NM}
\]
we get:
\[
\nonumber
\label{asymptotics:delta}
{\rm h}^{\lambda,n}_{c}
=
\lim\limits_{\Delta\to\infty}F^{\lambda, n}_{c, \Delta}
=
\sum_{|M|=|N|=n}  \left[ B^{\,{n}}_{c,\Delta}\right]_{NM}\left[ B^n_{c,\Delta} \right]^{MN}
=
\sum_{|N| = n} \delta_{N}^{N}
=
p(n)
\]
where $p(n)$ denotes the number of ways $n$ can be written as a sum of positive integers. Thus
\[
\nonumber
\label{limit:block:delta}
\lim_{\Delta\to\infty}\left(q^{\frac{c}{24}-\Delta}\mathcal{F}_{c, \Delta}^{\lambda}(q)\right)
= \prod\limits_{n=1}^{\infty}\left(1-q^n\right)^{-1}.
\]
This suggest the following definition of
the elliptic 1-point block on a torus:
\[
\nonumber \label{elliptic:definition} \mathcal{H}_{c,
\Delta}^{\lambda}(q) =
q^{\frac{c-1}{24}-\Delta}\,\eta(q)\,\mathcal{F}_{c,
\Delta}^{\lambda}(q) = \sum\limits_{n=0}^{\infty} q^n {H}_{c,
\Delta}^{\lambda,n}\ ,
\]
where $\eta(q)$ is the Dedekind eta function. One easily checks
that the coefficients ${H}_{c, \Delta}^{\lambda,n}$ have
essentially the same pole structure as the coefficients $
F^{\lambda, n}_{c, \Delta}$ and the following  recursive formula
holds:
\begin{eqnarray}
\label{elliptic:recursion:torus}
{H}_{c, \Delta}^{\lambda,n} & = &
\delta^n_0 + \hskip -5pt
\begin{array}[t]{c}
{\displaystyle\sum} \\[-6pt]
{\scriptscriptstyle
1 \leq rs \leq n}
\end{array}
\hskip -5pt \frac{ A_{rs}(c)\, P^{rs}_{c}\!\left[^{\hskip 10pt
\Delta_{\lambda}}_{\Delta_{rs}+rs}\right]
P^{rs}_{c}\!\left[^{\Delta_{\lambda}}_{\Delta_{rs}}\right] } {
\Delta-\Delta_{rs}(c) }\, {H}_{c, \Delta_{rs}+rs}^{\lambda,n-rs}\
.
\end{eqnarray}
Let us note that the form of the regular terms in (\ref{elliptic:recursion:torus})
can be also deduced from the limiting case $\Delta \to 0$ ($\lambda \to Q$) where
the fusion polynomials, and therefore all the residua, vanish
and the 1-point block become the Virasoro character.

\section{Poghossian identities}

Using (\ref{rho:explicit}) one can  check the identity:
\begin{eqnarray*}
P^{rs}_{c}\!\left[^{\hskip 10pt \Delta_{\lambda}}_{\Delta_{rs}+rs}\right]
P^{rs}_{c}\!\left[^{\Delta_{\lambda}}_{\Delta_{rs}}\right]
& = &
\!\!\!\!\!\!
\prod\limits_{\begin{array}{c}\\[-22pt]\scriptstyle p=1-r\\[-6pt]\scriptstyle p+r=1\,  {\rm mod}\, 2
\end{array}}^{r-1}
\!\!\!
\prod\limits_{\begin{array}{c}\\[-22pt]\scriptstyle q=1-s\\[-6pt]\scriptstyle q+s=1\, {\rm  mod}\, 2
\end{array}}^{s-1}
\!\!\!\!\!\!\!\!\!\left(\frac{\lambda}{2} + p b + q b^{-1} + \frac{Q}{2}\right)
\\
&& \hspace{-80pt}\times
\left(\frac{\lambda}{2} + p b + q b^{-1} - \frac{Q}{2}\right)
\left(\frac{\lambda}{2} + p b + q b^{-1} + \frac{b}{2} -  \frac{1}{2b}\right)
\left(\frac{\lambda}{2} + p b + q b^{-1} - \frac{b}{2} +  \frac{1}{2b} \right).
\end{eqnarray*}
The r.h.s. can be identified as a product of fusion polynomials:
\begin{eqnarray}
\label{rel:among:fusion:pol}
P^{rs}_{c}\!\left[^{\hskip 10pt \Delta_{\lambda}}_{\Delta_{rs}+rs}\right]
P^{rs}_{c}\!\left[^{\Delta_{\lambda}}_{\Delta_{rs}}\right]
& = &
(16)^{rs}\,P^{rs}_{c}\!\left[^{\Delta_2}_{\Delta_1}\right]\,
P^{rs}_{c}\!\left[^{\Delta_3}_{\Delta_4}\right]
\end{eqnarray}
provided that one of the equalities:
\begin{eqnarray}
\nonumber
(\lambda_1,\lambda_2, \lambda_3, \lambda_4)
& = &
\left(\frac{Q}{2},\frac{\lambda}{2}, \frac{\lambda}{2}, \frac{b}{2} - \frac{1}{2b}\right),
\\[6pt]
\label{parametrization:2}
( \lambda_1,\lambda_2,\lambda_3, \lambda_4)
& = &
\left( \frac{1}{2b},\frac{\lambda}{2}+\frac{b}{2},\frac{\lambda}{2}-\frac{b}{2}, \frac{1}{2b}\right),
\\[6pt]
\nonumber
(\lambda_1,\lambda_2, \lambda_3, \lambda_4)
& = &
\left(\frac{b}{2},\frac{\lambda}{2}+\frac{1}{2b},  \frac{\lambda}{2}- \frac{1}{2b}, \frac{b}{2}\right),
\end{eqnarray}
hold.
Let us now recall that in the case of the 4-point elliptic block on the sphere
\begin{eqnarray*}
\mathcal{H}_{c,\Delta}\left[^{\Delta_3\:\Delta_2}_{\Delta_4\:\Delta_1}\right]\!(q)
& = & 1 + \sum\limits_{n=1}^{\infty}
(16q)^n\,H^n_{c,\Delta}\left[^{\Delta_3\:\Delta_2}_{\Delta_4\:\Delta_1}\right]
\end{eqnarray*}
the recursive relation takes the form \cite{Zamolodchikov:2,Zamolodchikov:3}:
\begin{eqnarray}
\label{elliptic:recursion:sphere}
H^n_{c,\Delta}\left[^{\Delta_3\:\Delta_2}_{\Delta_4\:\Delta_1}\right]
& = &
\delta^n_0
+
\hskip -5pt
\sum\limits_{1 \leq rs \leq n}
\hskip -5pt
\frac{
A_{rs}(c)\,
P^{rs}_{c}\!\left[^{\Delta_2}_{\Delta_1}\right]\,
P^{rs}_{c}\!\left[^{\Delta_3}_{\Delta_4}\right]
}
{
\Delta-\Delta_{rs}(c)
}\,
H^{n-rs}_{c,\Delta_{rs} + rs}\left[^{\Delta_3\:\Delta_2}_{\Delta_4\:\Delta_1}\right].
\end{eqnarray}
Since the solutions to the recursive formulae  (\ref{elliptic:recursion:torus}) and (\ref{elliptic:recursion:sphere}) are unique
we get by comparing
(\ref{elliptic:recursion:torus}),  (\ref{rel:among:fusion:pol}) and (\ref{elliptic:recursion:sphere}) that
\[
{H}_{c, \Delta}^{\lambda,n} = 16^n
H^n_{c,\Delta}\left[^{\Delta_3\:\Delta_2}_{\Delta_4\:\Delta_1}\right]
\]
and therefore
\begin{equation}
\nonumber \label{rel:among:blocks} \mathcal{H}_{c,
\Delta}^{\lambda}(q) = \mathcal{H}_{c,\Delta}\left[^{\;\;\;{1\over
2}\lambda\;\;\;\;\;\;\;\;\;{1\over 2}\lambda}_{{{1\over
2}(b-{1\over b})\:\:{1\over 2}(b+{1\over b})}}\right]\!(q) =
\mathcal{H}_{c,\Delta}\left[_{\;\;\;\;\;\;{1\over
2b}\;\;\;\;\;\;\;\;\;{1\over 2b}}^{{{1\over 2}(\lambda -b
)\:\:{1\over 2}(\lambda +b)}}\right]\!(q) =
\mathcal{H}_{c,\Delta}\left[_{\;\;\;\;\;{1\over
2}b\;\;\;\;\;\;\;\;\;\;{1\over 2}b}^{{{1\over 2}(\lambda-{1\over
b})\:\:{1\over 2}(\lambda+{1\over b})}}\right]\!(q).
\end{equation}
We have thus obtained a simple proof of the first relation proposed  in \cite{Poghossian:2009mk}.

In the present notation the second relation conjectured in \cite{Poghossian:2009mk}  reads:
\begin{eqnarray}
\label{FLNOrelation} \nonumber
\mathcal{H}_{c,\Delta}\left[^{{1\over 2b}\;\;\mu}_{{{1\over
2b}\:\:{1\over 2b}}}\right]\!(q) &=&
\mathcal{H}_{c',\Delta'}\left[^{\;\;\;{1\over
\sqrt{2}}\mu\;\;\;\;\;\;{1\over \sqrt{2}}\mu}_{{{1\over
2}(b'-{1\over b'})\: \:{1\over 2}(b'+{1\over b'})}}\right]\!(q^2)
\\
&=& \mathcal{H}_{c',\Delta'}\left[_{\;\;\;\;\;\;{1\over
2b'}\;\;\;\;\;\;\;\;\;\;{1\over 2b'}}^{{{1\over \sqrt{2}}\mu
-{1\over 2}b' \;\;{1\over \sqrt{2}}\mu +{1\over
2}b'}}\right]\!(q^2)
\\
\nonumber &=& \mathcal{H}_{c',\Delta'}\left[_{\;\;\;\;\;\;{1\over
2}b'\;\;\;\;\;\;\;\;\;\;{1\over 2}b'}^{{{1\over \sqrt{2}}\mu
-{1\over 2b'} \;\;{1\over \sqrt{2}}\mu +{1\over
2b'}}}\right]\!(q^2).
\end{eqnarray}
where:
\begin{eqnarray}
\nonumber
\label{identification}
c'&=& 1+6\left(b'+{1\over b'}\right)^2
,\;\;\;\;\;\;
\Delta_{\lambda'}' \;=\; \frac{(b' + \frac{1}{b'})^2}{4}- \frac{(\lambda')^2}{4}\ ,
\\[-6pt]
\\[-6pt]
\nonumber
b'&=&\sqrt{2}b\ ,\;\;\;\;\;\; \lambda'=\frac{\lambda}{\sqrt{2}}\ .
\end{eqnarray}
There is also another relation of a similar origin with $b' = {b\over \sqrt{2}}$:
\begin{eqnarray}
\nonumber \label{FLNOrelationII}
\mathcal{H}_{c,\Delta}\left[^{{b\over 2}\;\;\mu}_{{{b\over
2}\:\:{b\over 2}}}\right]\!(q) &=&
\mathcal{H}_{c',\Delta'}\left[^{\;\;\;{1\over
\sqrt{2}}\mu\;\;\;\;\;\;{1\over \sqrt{2}}\mu}_{{{1\over
2}(b'-{1\over b'})\: \:{1\over 2}(b'+{1\over b'})}}\right]\!(q^2)
\\
&=& \mathcal{H}_{c',\Delta'}\left[_{\;\;\;\;\;\;{1\over
2b'}\;\;\;\;\;\;\;\;\;\;{1\over 2b'}}^{{{1\over \sqrt{2}}\mu
-{1\over 2}b' \;\;{1\over \sqrt{2}}\mu +{1\over
2}b'}}\right]\!(q^2)
\\
\nonumber &=& \mathcal{H}_{c',\Delta'}\left[_{\;\;\;\;\;\;{1\over
2}b'\;\;\;\;\;\;\;\;\;\;{1\over 2}b'}^{{{1\over \sqrt{2}}\mu
-{1\over 2b'} \;\;{1\over \sqrt{2}}\mu +{1\over
2b'}}}\right]\!(q^2).
\end{eqnarray}

We shall show that  relations of the form:
\begin{equation}
\nonumber
\label{HqHq2}
 \mathcal{H}_{c,\Delta}\left[^{ \eta \;  \mu}_{\eta \;\eta} \right](q)
=
 \mathcal{H}_{c', \Delta'}\left[^{ \lambda_3' \;  \lambda_2'}_{\lambda_4' \;\lambda_1'} \right](q^2)
\end{equation}
are to large extent unique. Let us first observe that the residua
of the coefficients of $\mathcal{H}_{c,\Delta}\left[^{ \eta \;
\mu}_{\eta \;\eta} \right](q)$ contain the fusion polynomial
\begin{eqnarray*}
P^{rs}_{c}\!\left[^{{\eta}}_{{\eta}}\right] \; =
\!\!\!\!\!\!
\prod\limits_{\begin{array}{c} \\[-22pt]\scriptstyle p=1-r\\[-6pt]\scriptstyle p+r=1\,  {\rm mod}\, 2
\end{array}}^{r-1}
\!\!\!
\prod\limits_{\begin{array}{c}\\[-22pt]\scriptstyle q=1-s\\[-6pt]\scriptstyle q+s=1\, {\rm  mod}\, 2
\end{array}}^{s-1}
\!\!\!\!\!\!\!\!\!
\left(\frac{2 \eta +p b + q b^{-1}}{2}\right) \left(\frac{p b + q b^{-1}}{2}\right)
\end{eqnarray*}
which always vanishes if both $r$ and $s$ are odd. Moreover if
$\eta = {1\over 2b}$ it vanishes for odd $r$ and all
$s$.\footnote{The case  $\eta = {b\over 2}$ leading to the
relation (\ref{FLNOrelationII}) can be analyzed in a similar way.}
Since $H^1_{c,\Delta}\left[^{ \eta \;  \mu}_{\eta \;\eta} \right]
=0$, it follows that for $\eta = {1\over 2b}$ all the odd
coefficients of $\mathcal{H}_{c,\Delta}\left[^{ \eta \;
\mu}_{\eta \;\eta} \right](q)$ vanish and the even ones satisfy
the recursive relation:
\begin{eqnarray}
\nonumber
\label{rek2}
H^{2m}_{c, \Delta} \left[^{ \eta \;  \mu}_{\eta \;\eta} \right]
=\delta^n_0 +
 \hspace{-10pt}
\begin{array}[t]{c}
{\displaystyle\sum} \\[-6pt]
{\scriptscriptstyle
k,s \in \mathbb{N}}
 \\[-8pt] {\scriptscriptstyle 1 \leq ks \leq m}
\end{array}
 \frac{R^{2k,s}_{c}\!\left[^{ \eta \;  \mu}_{\eta \;\eta} \right]}{\Delta - \Delta_{2k,s}} \,
H^{2m-2ks}_{c,\Delta_{2k,s}+2ks}\!\left[^{ \eta \;  \mu}_{\eta \;\eta} \right].
\end{eqnarray}
This is to be compared with the recursive relation:
\begin{eqnarray}
\nonumber
\label{rek3}
H^m_{b', \Delta'}\left[^{ \lambda_3' \;  \lambda_2'}_{\lambda_4' \;\lambda_1'} \right]
= \delta^n_0+\hspace{-10pt}
\begin{array}[t]{c}
{\displaystyle\sum} \\[-6pt]
{\scriptscriptstyle
k,s\in \mathbb{N}}
 \\[-8pt] {\scriptscriptstyle 1 \leq ks \leq m}
\end{array}
 \frac{R^{k,s}_{c'}\!\left[^{ \lambda_3' \;  \lambda_2'}_{\lambda_4' \;\lambda_1'} \right]}{\Delta' - \Delta_{rs}'} \,
H^{m-ks}_{b',\Delta_{k,s}'+ks}\!\left[^{ \lambda_3' \;  \lambda_2'}_{\lambda_4' \;\lambda_1'} \right].
\end{eqnarray}
The numbers of terms in both relations coincide. Moreover for
the identification of parameters (\ref{identification}):
\begin{equation}\label{bieguny2ks}
(\Delta-\Delta_{2k,s}) = 2 \left[ \frac{1}{4} (\sqrt{2}kb + \frac{s}{\sqrt{2}b})^2 - \frac{\lambda^2}{8}  \right]
= 2 (\Delta'-\Delta_{k,s}'),
\end{equation}
and:
\begin{equation}\label{Drs+rs}
 \Delta_{2ks} + 2ks = \Delta_{ks}' + ks .
\end{equation}
The fusion polynomials can be expressed in the form:
\begin{eqnarray*}
P^{2k,s}_{c}\!\left[^{\frac{1}{2b}}_{\frac{1}{2b}}\right]
P^{2k,s}_{c}\!\left[^{\;\mu}_{\frac{1}{2b}}\right]
&=&
4^{-ks}
\!\!\!\!\!\!
\prod\limits_{\begin{array}{c} \\[-22pt]\scriptstyle p=1-2k\\[-6pt]\scriptstyle p+2k=1\,  {\rm mod}\, 2
\end{array}}^{2k-1}
\!\!\!
\prod\limits_{\begin{array}{c}\\[-22pt]\scriptstyle q=1-s\\[-6pt]\scriptstyle q+s=1\, {\rm  mod}\, 2
\end{array}}^{s-1}
\!\!\!\!\!\!\!\!\!
\left(\frac{p b + (q+1) b^{-1}}{2}\right) \left(\frac{p b + q b^{-1}}{2}\right)
\\ \nonumber
&& \hspace{-70pt}\times
\!\!\!\!\!\!
\prod\limits_{\begin{array}{c} \\[-22pt]\scriptstyle p=1-kr\\[-6pt]\scriptstyle p+k=1\,  {\rm mod}\, 2
\end{array}}^{k-1}
\!\!\!
\prod\limits_{\begin{array}{c}\\[-22pt]\scriptstyle q=1-s\\[-6pt]\scriptstyle q+s=1\, {\rm  mod}\, 2
\end{array}}^{s-1}
\!\!\!\!\!\!\!\!\!
\hspace{-5pt}
\Bigg(\left(\frac{\mu}{\sqrt{2}} + \frac{1}{2b'} - \frac{b'}{2} + p b' + \frac{q}{b'}\right)
\left(\frac{\mu}{\sqrt{2}}- \frac{1}{2b'} - \frac{b'}{2} + p b' + \frac{q}{b'}\right)
\\ \nonumber
& & \hspace{30pt}
\left(\frac{\mu}{\sqrt{2}}+\frac{1}{2b'} + \frac{b'}{2} + p b' + \frac{q}{b'}\right)
\left(\frac{\mu}{\sqrt{2}}-\frac{1}{2b'} + \frac{b'}{2} + p b' + \frac{q}{b'}\right) \Bigg).
\end{eqnarray*}
As in the case of our previous derivation the last two lines can be identified as the fusion polynomials:
\begin{eqnarray}
\label{fuz2ks}
\nonumber
P^{2k,s}_{c}\!\left[^{\frac{1}{2b}}_{\frac{1}{2b}}\right]
P^{2k,s}_{c}\!\left[^{\;\mu}_{\frac{1}{2b}}\right]
&=&
4^{-ks}
\!\!\!\!\!\!
\prod\limits_{\begin{array}{c} \\[-22pt]\scriptstyle p=1-2k\\[-6pt]\scriptstyle p+2k=1\,  {\rm mod}\, 2
\end{array}}^{2k-1}
\!\!\!
\prod\limits_{\begin{array}{c}\\[-22pt]\scriptstyle q=1-s\\[-6pt]\scriptstyle q+s=1\, {\rm  mod}\, 2
\end{array}}^{s-1}
\!\!\!\!\!\!\!\!\!
\left(\frac{p b + (q+1) b^{-1}}{2}\right) \left(\frac{p b + q b^{-1}}{2}\right)
\\[-6pt]
\\[-6pt]
\nonumber
&\times&
(16)^{ks}\,P^{ks}_{c'}\!\left[^{\lambda_2'}_{\lambda_1'}\right]\,
P^{ks}_{c'}\!\left[^{\lambda_3'}_{\lambda_4'}\right]
\nonumber
\end{eqnarray}
where one of the following choices is assumed:
\begin{eqnarray}
\nonumber
(\lambda_1',\lambda_2', \lambda_3', \lambda_4')
& = &
\left(\frac{\mu}{\sqrt{2}}, \frac{b'}{2}+\frac{1}{2b'},\frac{\mu}{\sqrt{2}}, \frac{b'}{2} - \frac{1}{2b'}\right),
\\[6pt]
\label{parametrization:2a}
\left(\lambda_1',\lambda_2', \lambda_3', \lambda_4'\right)
& = &
\left(\frac{\mu}{\sqrt{2}}+\frac{b'}{2}, \frac{1}{2b'},\frac{\mu}{\sqrt{2}}-\frac{b'}{2}, \frac{1}{2b'}\right),
\\[6pt]
\nonumber
\left(\lambda_1',\lambda_2', \lambda_3', \lambda_4'\right)
& = &
\left(\frac{\mu}{\sqrt{2}}+\frac{1}{2b'}, \frac{b'}{2}, \frac{\mu}{\sqrt{2}}- \frac{1}{2b'}, \frac{b'}{2}\right).
\end{eqnarray}
Finally calculating the coefficients
$A_{rs}$ one gets:
\begin{equation}
\label{A2ks}
A_{2k,s}(c)
=
2^{1-2ks}\,A_{ks}(c')
\hskip -10pt
\prod\limits_{\begin{array}{c} \\[-22pt]\scriptstyle p=1-2k\\[-6pt]\scriptstyle p+2k=1\,  {\rm mod}\, 2
\end{array}}^{2k-1}
\!\!\!
\prod\limits_{\begin{array}{c}\\[-22pt]\scriptstyle q=1-s\\[-6pt]\scriptstyle q+s=1\, {\rm  mod}\, 2
\end{array}}^{s-1}
\!\!\!\!\!\!\!\!\!
\left(p b + (q+1) b^{-1}\right)^{-1} \left(p b + q b^{-1}\right)^{-1},
\end{equation}
where:
$$
A_{ks}(c')= \frac12
\hspace{-35pt}\prod_{\begin{array}{c}\\[-24.5pt]\scriptstyle p=1-k\\[-6pt]\hspace{30pt}\scriptstyle (p,q)\neq (0,0),(k,s)
\end{array}}^{k}
\hspace{-35pt}\prod_{q=1-s}^{s}(p b' + \frac{q}{b'})^{-1}.
$$
Taking into account formulae
(\ref{bieguny2ks}), (\ref{Drs+rs}), (\ref{fuz2ks}) and (\ref{A2ks}) one obtains:
$$
\frac{R^{2k,s}_{c}\!\left[^{ {1\over 2b} \; \; \mu}_{{1\over 2b} \;{1\over 2b}} \right]}{\Delta - \Delta_{rs}}
= 16^{-ks} \frac{ R^{k,s}_{c'}\!\left[^{ \lambda_3' \;  \lambda_2'}_{\lambda_4' \;\lambda_1'} \right]}
 {(\Delta' - \Delta_{rs}')}\ .
$$
Hence the coefficients
$
H^{2m}_{c, \Delta} \left[^{ \eta \;  \mu}_{\eta \;\eta} \right]$ and $
(16)^{-m}  H^m_{b', \Delta'}\left[^{ \lambda_3' \;  \lambda_2'}_{\lambda_4' \;\lambda_1'} \right]
$
with the weights (\ref{parametrization:2a}) satisfy the same recursive relations. This completes our proof of
the formula (\ref{FLNOrelation}).
Formula (\ref{FLNOrelationII}) can be derived along the same lines starting with $\eta = {b\over 2}$.

\section{Modular bootstrap in Liouville theory}

The modular invariance of the 1-point function on the torus:
\begin{equation}
\label{modinv}
\langle \phi_{\lambda,\bar \lambda} \rangle_{-{1\over \tau}} =
(-1)^{\Delta_\lambda-\Delta_{\bar\lambda}}\tau^{\Delta_\lambda}
\bar\tau^{\Delta_{\bar\lambda}}\langle \phi_{\lambda,\bar \lambda} \rangle_\tau
\end{equation}
along with the crossing invariance of the 4-point function on the sphere form
the basic consistency conditions for any CFT on closed surfaces \cite{Sonoda:1988fq}.
 In  the case of the  Liouville theory the 1-point function can be expressed in terms of the elliptic blocks
 as follows:
$$
\langle \phi_{\lambda} \rangle_{\tau} = \int\limits_{i\mathbb{R}_+}
{\mathrm{d}\alpha  \over 2i}\left| q^{\Delta_\alpha - \frac{Q^2}{4}} \, \eta(q)^{-1}
\mathcal{H}_{c, \Delta_\alpha}^{\lambda}(q) \right|^2 \,
C^{\lambda}_{\Delta_\alpha}\ ,
 $$
 where $q={\rm e}^{2\pi i \tau}$.
Condition (\ref{modinv}) then takes the form:
\begin{equation}
\label{modinvlio}
\int\limits_{i\mathbb{R}_+} {\mathrm{d}\alpha  \over 2i}
\left| \tilde q^{-{\alpha^2\over 4}}
\mathcal{H}_{c, \Delta_\alpha}^{\lambda}(\tilde q) \right|^2 \, C^{\lambda}_{\Delta_\alpha}
=
|\tau|^{2 \Delta_{\lambda}+1}
\int\limits_{i\mathbb{R}} {\mathrm{d}\alpha  \over 2i}
\left| q^{-{\alpha^2\over 4}}
\mathcal{H}_{c, \Delta_\alpha}^{\lambda}(q) \right|^2 \, C^{\lambda}_{\Delta_\alpha}
\end{equation}
where
and $\tilde q= {\rm e}^{-2\pi i {1\over \tau}}$.
The Liouville structure constant reads \cite{Zamolodchikov:1995aa}:
$$
 C^{\lambda}_{\Delta_\alpha} =
\left[ \pi \mu \gamma(b^2) b^{2-2b^2}\right]^{-\frac{Q+\lambda}{2b}} \times
\frac{\Upsilon_0 \Upsilon(Q + \lambda) \Upsilon(\alpha) \Upsilon( -\alpha)}
{ \Upsilon^2\left(\frac{Q}{2} + {\lambda\over 2}\right) \Upsilon\left(\frac{Q}{2} + {\lambda\over 2} + \alpha\right)
 \Upsilon\left(\frac{Q}{2} + {\lambda\over 2} - \alpha\right)}\ .
$$
Separating the $\alpha$-dependent part
\begin{eqnarray}
\nonumber
C^{\lambda}_{\Delta_\alpha}  &=&  4 \left[ \pi \mu \gamma(b^2) b^{2-2b^2}\right]^{-\frac{Q+\lambda}{2b}} \times
\frac{\Upsilon_0 \Upsilon(Q + \lambda)}{\Upsilon^2\left(\frac{Q}{2} + {\lambda\over 2}\right)} \, r(\alpha)
\\[-10pt]
\nonumber
\\[0pt]
\nonumber
\label{r}
r(\alpha) &=&- {\alpha^2\over 4} \exp{\int_0^\infty {\mathrm{d}t\over t} \Bigg( \frac{1 + b^4 - b^2 (\lambda^2+2)}{2 b^2 } e^{-t}
+ \cosh(\alpha t)\, \frac{\cosh\left({t \lambda\over 2}\right) - \cosh\left(\frac{( b^2-1) t}{2 b}\right)}{\sinh(\frac{b t}{2}) \sinh(\frac{t}{2 b})}
 \Bigg)}
\end{eqnarray}
one can write (\ref{modinvlio}) as:
\begin{equation}
\label{numerical}
\int\limits_{i\mathbb{R}} {\mathrm{d}\alpha  \over 2i}
\left| \tilde q^{-{\alpha^2\over 4}} \,
\mathcal{H}_{c, \Delta_\alpha}^{\lambda}(\tilde q) \right|^2 \, r(\alpha)
=
|\tau|^{2 \Delta_{\lambda}+1}
\int\limits_{i\mathbb{R}} {\mathrm{d}\alpha  \over 2i}
\left| q^{-{\alpha^2\over 4}} \,
\mathcal{H}_{c, \Delta_\alpha}^{\lambda}(q) \right|^2 \, r(\alpha)\ .
\end{equation}
This relation can be numerically analyzed with the help of the recursion relations derived in Section~3.
Due to the rapidly oscillating integrand in (\ref{r}) the numerical
calculation of function $r(t)$ has to be carefully done. We present a sample of the calculations
for $c=2, \lambda =i$ and for the modular parameter $\tau$ along the imaginary axis
in the range $[0.2 i, 5i]$.
The results for the elliptic block expanded up to the term $q^n$, $n=4,5,6$
are presented  on Fig.1. where the relative difference of the left and the right side
of (\ref{numerical}) is plotted.

\begin{figure}[ht]
\centering
\includegraphics{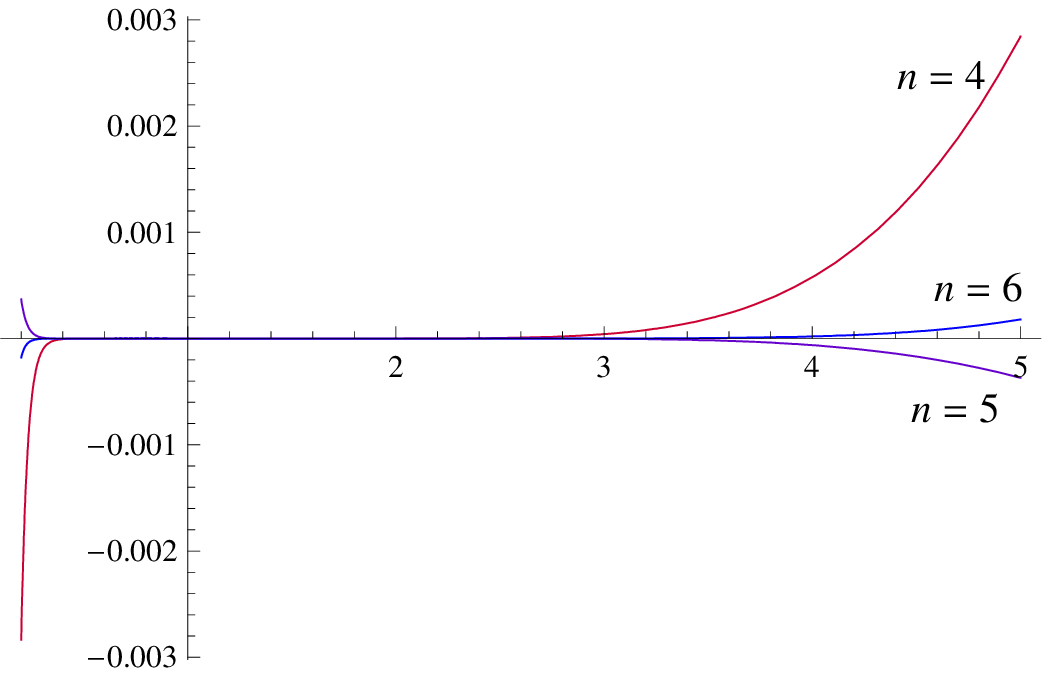}

\vskip 5pt
Fig.1 Numerical check of the modular bootstrap.
\end{figure}

\setcounter{equation}{0}
\section*{Acknowledgements}
This work  was  supported by the Polish State Research
Committee (KBN) grant no. N N202 0859 33.
The work of L.H. was also supported by MNII grant 189/6.PRUE/2007/7.

\end{document}